\documentclass[9pt, reqno]{amsart}

\usepackage[a4paper,twoside,left=30mm,right=30mm,top=40mm,bottom=35mm]{geometry}
\usepackage{amsmath}                
\usepackage{amsthm}                
\usepackage{amssymb}
\usepackage[foot]{amsaddr} 
\usepackage{multirow}                                                                
\usepackage{graphicx}      
\usepackage[usenames, dvipsnames,svgnames,table]{xcolor}
\usepackage[small]{caption}                                                                                               
\usepackage[figuresleft]{rotating}                     
\usepackage{setspace} 
\definecolor{myGreen}{HTML}{7d9b76}		
\definecolor{myBrown}{HTML}{4a3139}		
\definecolor{myBlack}{HTML}{2B2B2B}			                     
\usepackage[hidelinks, colorlinks=True, urlcolor=myGreen, citecolor=myBrown, linkcolor=myBrown]{hyperref}                             
\usepackage{subcaption}
\usepackage{physics}
\usepackage{mathtools}
\usepackage{booktabs}
\usepackage{arev}
\usepackage[T1]{fontenc}
\usepackage{floatrow}
\usepackage{svg}
\usepackage{enumitem}
\usepackage{arydshln}	
\usepackage{etoolbox} 
\usepackage{titlesec}
\usepackage[title,toc]{appendix}
\usepackage{abstract}
\usepackage[numbers]{natbib}
\usepackage{orcidlink}
\usepackage{soul}

\numberwithin{equation}{section}								
\newcommand{\boxtitle}[1]{
	\colorbox{myBrown!50}{%
		\color{myBlack}
		\transparent{1} %
		\enspace \begin{minipage}[c]{\linewidth-5.4\fboxsep}
			\vspace*{.25em} #1 %
		\end{minipage}%
	}
}
\titleformat{\section}{\normalfont\Large\bfseries\boxtitle}{\thesection}{2em}{\raggedright}
\titlespacing*{\section}{0em}{1em}{1em}
\newcommand{\boxsubtitle}[1]{
    \colorbox{myGreen!50}{%
		\color{myBlack}
		\transparent{1} %
		\enspace \begin{minipage}[c]{\linewidth-4.85\fboxsep}
			\vspace*{.25em} #1 %
		\end{minipage}%
	}
}
\titleformat{\subsection}{\normalfont\bfseries\boxsubtitle}{\thesubsection}{2em}{\raggedright}
\titlespacing*{\subsection}{0em}{1em}{1em}

\titleformat{\subsubsection}[runin]{\normalfont\itshape}{\thesubsubsection}{2em}{}[.]
\titlespacing*{\subsubsection}{0pt}{1em}{1em}

\makeatletter
\renewcommand{\@biblabel}[1]{[#1]\hfill}
\makeatother

\makeatletter
\patchcmd{\@maketitle}{\normalsize}{\Large}{}{} 
\makeatother    

\begin{document} 
	\title[Reaction-diffusion models of invasive tree pest spread]{Reaction-diffusion models of invasive tree pest spread: quantifying the expansion of oak processionary moth in the UK}

	\author[JP McKeown]{Jamie P McKeown\textsuperscript{C,1}\orcidlink{0009-0003-8130-5737}}
    \email{j.p.mckeown2@ncl.ac.uk}
	\author[LE Wadkin]{Laura E Wadkin\textsuperscript{1}\orcidlink{0000-0001-7355-2023}}
    \email{laura.wadkin@newcastle.ac.uk}
	\author[NG Parker]{Nick G Parker\textsuperscript{1}\orcidlink{0000-0002-7238-8492}} 
    \email{nick.parker@newcastle.ac.uk}
	\author[A Golightly]{Andrew Golightly\textsuperscript{2}\orcidlink{0000-0001-6730-1279
}} 
    \email{andrew.golightly@durham.ac.uk}
	\author[AW Baggaley]{Andrew W Baggaley\textsuperscript{1,3}\orcidlink{0000-0002-6813-2443}}
    \email{a.baggaley1@lancaster.ac.uk}
	
    \address[C]{Corresponding author: \href{mailto:j.p.mckeown2@newcastle.ac.uk}{j.p.mckeown2@newcastle.ac.uk}}
	\address[1]{School of Mathematics, Statistics, and Physics, Newcastle University, Newcastle upon Tyne, UK.}
	\address[2]{Department of Mathematical Sciences, Durham University, Durham, UK.}
	\address[3]{School of Mathematical Sciences, Lancaster University, Lancaster, UK.}
	
	\date{\today}
	
	\maketitle

	\begin{abstract}
	\centering \bigskip
	 \begin{minipage}{\dimexpr\paperwidth-10cm}
		UK woodlands, forests, and urban treescapes are under threat from invasive species, exacerbated by climate change, trade, and transport. Invasive tree pests debilitate their host and disrupt forest ecosystems, thus it is imperative to quantitatively model and predict their spread. Addressing this, we model the spread of an invasive pest using a spatiotemporal reaction-diffusion equation, representing the spatial distribution as a population density field. We solve this intractable equation numerically and, from the solution, we determine first arrival times of the pest at locations in the field. The adopted model permits us to obtain the expansion rate of pest spread directly from the model parameters, which we infer in the Bayesian paradigm, using a Markov chain Monte Carlo scheme. We apply our framework to the ongoing spread of oak processionary moth in the UK, an outbreak which continues to grow despite management efforts. We demonstrate that our approach effectively captures the spread of the pest and that this has occurred at a non-constant expansion rate. The proposed framework is a powerful tool for quantitatively modelling the spread of an invasive tree pest and could underpin future prediction and management approaches.  \\
		
		
	\end{minipage} 
	\end{abstract}

\section{Introduction}\label{sec:introduction}
Invasive tree pests pose a significant ecological, economical, and epidemiological threat both in the UK and globally \cite{eschen_updated_2023, kenis_ecological_2009}. The UK hosts at least 121 pests of native tree species which were either introduced or have uncertain origin \cite{downey_state_2025}, and management of the most expensive six is estimated to cost around £919.9 million per year \cite{eschen_updated_2023}. Climate change is anticipated to favour successful establishment of satellite populations and international trade is known to be a key driver of these introductions \cite{downey_state_2025, Potter2017}. In particular, imports of live oak led to the successful establishment of the oak processionary moth (OPM) in the South of England in 2006 \cite{mindlin_arrival_2012}. OPM caterpillars -- which defoliate and weaken trees -- develop urticating setae which can shed and become airborne \cite{blaser_oak_2022}. Human and/or animal contact with the caterpillars, their detached fibrous hairs, or their nests' detritus can lead to cases of caterpillar dermatitis \cite{blaser_oak_2022, maier_caterpillar_2004, tomlinson_managing_2015}. Inhalation of the setae, which can break apart into microscopic fibres, can lead to respiratory irritation \cite{tomlinson_managing_2015}. \\ 

Since its establishment in the UK, the oak processionary moth has continued to expand its range, from its original introduction in South-West London \cite{suprunenko_estimating_2021}. There is a separate finding further afield in the Midlands, which is under eradication measures \cite{Hoppit2025}. Environmental factors, such as temperature, precipitation, and host species distribution, are expected to play a key role in successful colony establishment \cite{csoka_weather-dependent_2018, sands_population_2017}. Specifically, the predicted drier and hotter UK summers align well with both the OPM reproductive cycle and their preference for drought-like conditions \cite{csoka_weather-dependent_2018}. The climate in South England is already considered highly climatically-suitable for OPM, and Northern England and Scotland are expected to become highly suitable for OPM establishment by $2050$ to $2070$ \cite{godefroid_current_2020}. Thus, there is urgency for a robust framework for modelling invasive tree pests and OPM in particular, to predict expansion and inform the government's ongoing surveillance and management programme \cite{tomlinson_managing_2015}. \\

Oak processionary moth is a species of the order \textit{Lepidoptera}, undergoing complete metamorphosis during its one-year development cycle \cite{blaser_oak_2022}. The grey-white moths emerge between July and September, dispersing, mating, and ovipositing during an adult phase lasting 3 to 5 days \cite{blaser_oak_2022}. The collective population reproduces and disperses in a single annual event \cite{blaser_oak_2022}; any new nests in a given year are established by individuals who dispersed in the previous year. Estimates of the dispersal capabilities of adult OPM have been given as upwards of $20$km, however this figure is for males \cite{stigter_thaumetopoea_1997}. Female OPM seldom nest far from their parent nest \cite{blaser_oak_2022, sands_population_2017} and new nests are typically found within an average distance of $0.5$km of the parent nest \cite{sands_population_2017}. \\ 

Recent efforts to quantify the spread of OPM in the UK have included the following. Townsend \textit{et al} used amalgamated survey data from 2006 to 2012 to estimate the expansion rate of OPM in the UK \cite{townsend_oak_2013}. The same period of expansion was modelled using electric network theory by Cowley \textit{et al} \cite{cowley_using_2015}. In both cases, this (mostly) precedes the management strategy adopted in 2011, as well as the ongoing improvement to survey methodologies \cite{suprunenko_estimating_2021, wadkin_quantifying_2023}. Suprunenko \textit{et al} later determined expansion rate estimates up to 2019, applying a maximum distance method directly to the survey data \cite{suprunenko_estimating_2021}. They characterised OPM spread by two phases of expansion: 1) a slower phase, with population expansion at a rate of 1.66km/year from 2006 to 2014, followed by 2) a faster phase, at 6.17km/year, from 2015 to 2019 \cite{suprunenko_estimating_2021}. The authors suggested two potential explanations for this apparent biphasic expansion: `stratified diffusion' (OPM may be capable of both short- and long-distance dispersal) and external factors, such as environmental heterogeneity, a reduction of active control, and annual variation in climatic conditions \cite{suprunenko_estimating_2021}. A network approach was adopted in \cite{wadkin_quantifying_2023, golightly_accelerating_2023}, with which the authors demonstrated the significance of kilometre-scale population dynamics on the population growth in Richmond and Bushy Parks -- `hot spots' for OPM presence in the UK. This extended a previous characterisation of the population dynamics of OPM, which employed a stochastic susceptible-infested-removed model to estimate the time-varying infestation rate in the aforementioned parks \cite{wadkin_inference_2022}. Such compartmental models are often employed to study pest and pathogen spread, accounting for both reproductive and dispersal dynamics \cite{wadkin_inference_2022, suprunenko_predicting_2025}. \\

As outlined above, there is a pressing need to develop approaches to model and quantify the spread of OPM and invasive pests more generally. Motivated by this, here we develop an approach to model the spread of OPM based on a reaction-diffusion equation with parameters inferred from observational data. Reaction-diffusion equations have been applied to model populations of pine processionary moths \cite{robinet_potential_2014}, mosquitoes \cite{takahashi_mathematical_2005}, plant-pathogen spread \cite{Leclerc2023}, and Neolithic humans \cite{baggaley_bayesian_2012}, as well as the recent spread of COVID-19 \cite{viguerie_diffusionreaction_2020}. Here, we represent the spatial distribution of the pest as a population density field which evolves according to a spatiotemporal reaction-diffusion equation. Adopting a fixed density threshold, we apply a binary mask to this density field to obtain a presence/absence field for OPM infestation. Consequently, we can determine (simulated) arrival times of OPM under a given parameter configuration. We adopt a Bayesian approach to inferring model parameters, employing an adaptive Markov chain Monte Carlo (MCMC) scheme to estimate the expansion rate of OPM in the UK \cite{gamerman_markov_2006}. The aims of this paper are threefold: (1) demonstrate the viability of a reaction-diffusion model of tree pest spread (and OPM in particular); (2) determine an estimate of the current expansion rate of OPM in the UK, and; (3) challenge the proposed evidence that the expansion of OPM in the UK occurs at a non-constant rate. \\

\section{Materials and Methods}\label{sec:methods}
\begin{center}
    \begin{figure}[!ht]
        \centering 
        \includegraphics[width=\linewidth]{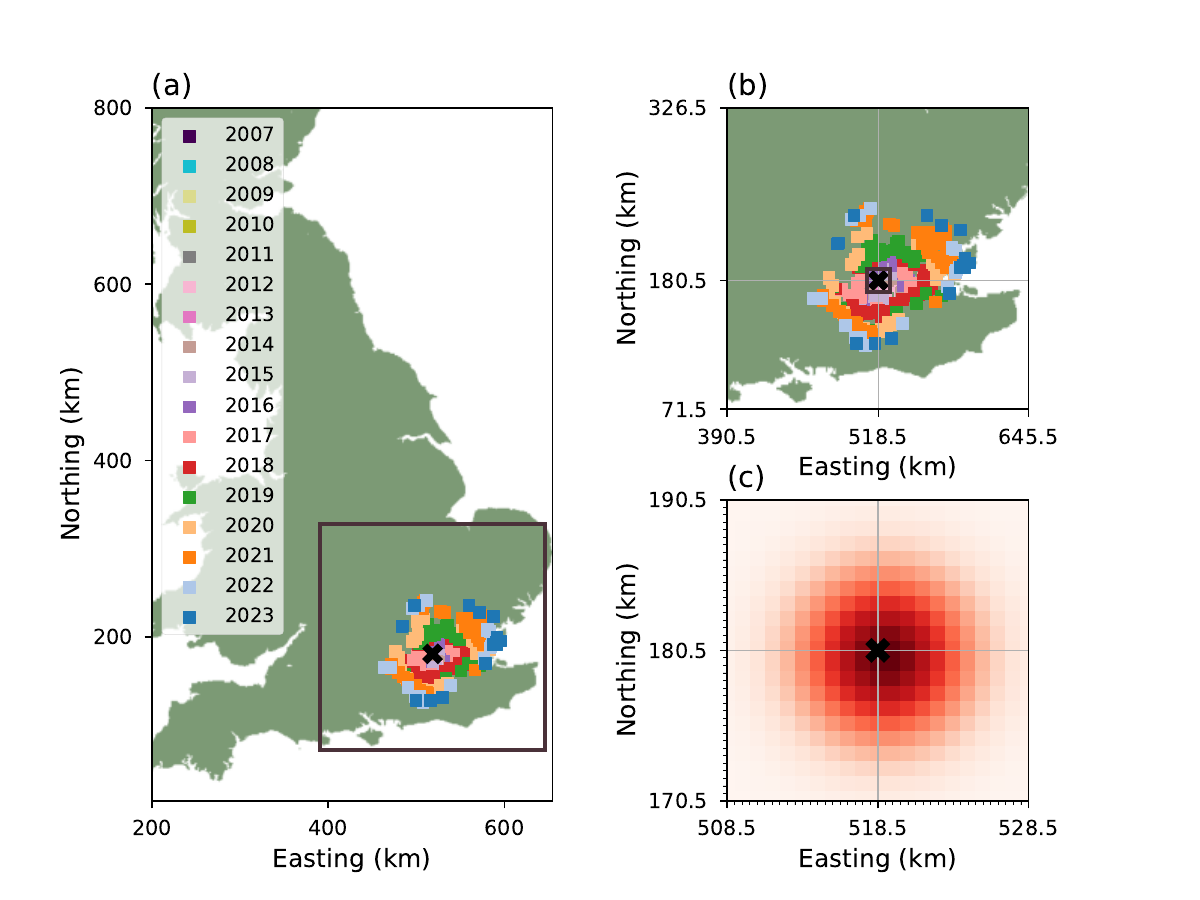}
        \caption{The distribution of OPM nest observations (see Section \ref{sec:data} for details). Each of the $414$ coloured squares indicates a $1\text{km}^2$ area where OPM is observed; the colour indicates the earliest year OPM was observed in that area. Black crosses indicate the centre of the initial distribution. Panel (a) shows the distribution on a map of the UK. Panel (b) shows the $256\text{km}^2$ grid $\Omega$ adopted as the modelling domain (see Section \ref{sec:data}), indicated by a square outline on panel (a). Panel (c) shows the approximation of the initial (observed in 2006) distribution. The minor ticks indicate the $1$km resolution grid. For the background maps of the UK, we use the shapefile for Q. robur (English oak) provided by \cite{hill_abundance_2017} (available at \cite{hill_abundance_dataset}), normalised to one.}
        \label{fig:observeddata}
    \end{figure}
    \end{center}

Here we describe the adopted approach of this paper. In Section~\ref{sec:data}, we present the OPM observational data used to train our reaction-diffusion models; these observed sites are shown in Figure~\ref{fig:observeddata}. In Section~\ref{sec:model}, we describe the reaction-diffusion model of OPM spread -- the FKPP equation -- along with our assumptions on the population. In Section~\ref{sec:proposedmodels}, we define the two models of OPM spread proposed in this work. In Section~\ref{sec:bayesianinference}, we outline the adopted Bayesian approach to inferring the parameters of these proposed models. Finally, in Section~\ref{sec:modelval}, we provide the metrics used to validate these models. 

\subsection{OPM observational data}\label{sec:data}	
\begin{figure}[!ht]
        \centering 
        \includegraphics[width=.8\linewidth]{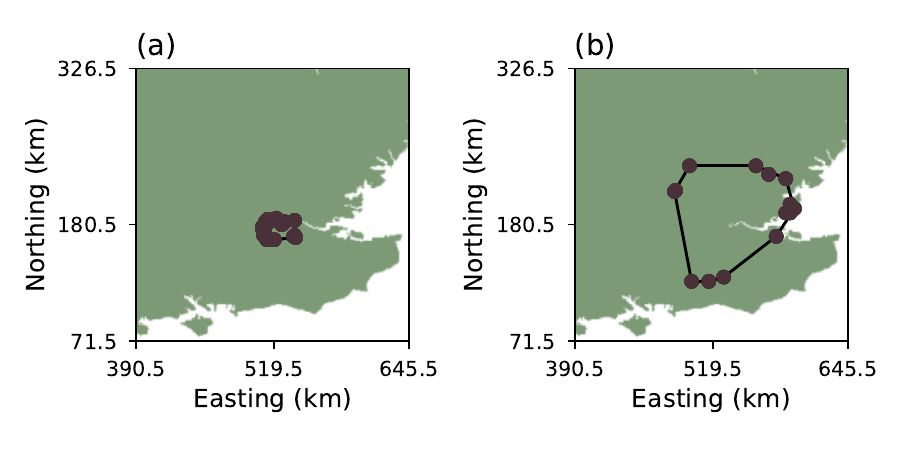}

    \caption{The distributions of observed sites for 2014 (panel (a)) and 2023 (panel (b)), obtained by applying a convex layer decomposition (Section~\ref{sec:data}). For the background maps of the UK, we use the shapefile for Q. robur (English oak) provided by \cite{hill_abundance_2017}, (available at \cite{hill_abundance_dataset}) normalised to one.}
    \label{fig:convexlayers}	
    \end{figure}

These data are derived from amalgamated survey data (provided by the Forestry Commission and the University of Southampton), obtained from yearly surveys conducted between $2006$ and $2023$ \cite{Branson2025}. The adopted reaction-diffusion model (see Section~\ref{sec:model}) captures the first arrival of OPM, but cannot capture the complex spatiotemporal processes behind the wavefront. Thus, we thin the dataset to capture this observed first arrival using a convex layer decomposition \cite{Yalcin2017}. Beginning with the final survey year $t = 2023$, we calculate the convex hull of (the locations of) the corresponding observed sites and interpret the vertex set of the hull as the observed wavefront in year $t$. We compare the sites observed in \emph{previous} years to the observed wavefront and remove any that are located \emph{ahead} of the wavefront. We set $t \gets t-1$ and repeat until (and including) the initial year $t=2006$. Then, beginning at $t=2006$, we construct the observed wavefront and compare this wavefront to sites observed in \emph{later} years, removing those located \emph{behind} the wavefront. We set $t \gets t+1$ and repeat until (and including) $t=2023$. This produces `layered' data that captures the spatial trend of population spread over time -- see Figure~\ref{fig:convexlayers} for the resulting observed distributions for $2014$ and $2023$. 

The observed data are point locations where OPM was first observed between $2006$ and $2023$, mapped to $414$ $1\text{km}^2$ areas (see Figure~\ref{fig:observeddata}). These locations are provided as six-digit easting-northing Ordnance Survey National Grid coordinates which have units of metres. The areas are located within a $256^2\text{km}^2$ square region $\Omega$ (panel (b) of Figure~\ref{fig:observeddata}). We discretize $\Omega$ into a $\Delta \mathbf{x} = 1$km-resolution grid such that each $\mathbf{x}_{ij} \in \Omega$ is given by $\mathbf{x}_{ij} = (e_j, n_i)$, where $e_j$ (resp. $n_i$) is the easting (resp. northing) coordinate of the centre of the corresponding $1\text{km}^2$ area, with units of km. Thus, each of the $414$ OPM observations consists of a location $\mathbf{x}_{ij}$ and a survey year $t_k$ when OPM was first observed in the $1\text{km}^2$ area with centre $\mathbf{x}_{ij}$. We refer to the pair $\mathbf{s}_k = (\mathbf{x}_{ij}, t_k)$ as the $k^\text{th}$ \emph{observed site}, where $k=1,\dots,414$. Note that the region $\Omega$ is large (compared to the extent of the observed data) to avoid boundary effects in our spatial simulations. \\

The initial (observed in $2006$) sites corresponded to five observations scattered over a small region. For our reaction-diffusion simulations, we take the initial condition of the population density field to be a Gaussian profile centred on and encapsulating these sites. We provide a plot of the initial distribution in panel (c) of Figure~\ref{fig:observeddata}. The centre of the initial distribution is $\mathbf{x}^c = (518.5,180.5)$km, indicated by a black cross on Figure~\ref{fig:observeddata}.\\

\subsection{FKPP equation for tree pest spread}\label{sec:model} 
The FKPP equation describes the evolution of a population density $N(\mathbf{x},t)$ at position $\mathbf{x}$ and time $t$, given by 
\begin{equation}
	\pdv{N}{t} = D\nabla^2 N + rN\left(1 - \frac{N}{K}\right),
	\label{eq:fkpp}
\end{equation}
with growth rate $r$, diffusivity $D$, and carrying capacity $K$ \cite{fisher_wave_1937}. The first term in Equation~\eqref{eq:fkpp} describes diffusion of $N$ in space; diffusion represents dispersal of the population. The second term describes the logistic growth of $N$ in time, where $r$ is the growth rate and $K$ the carrying capacity; logistic growth represents reproduction of the population subject to resource constraints encoded by $K$, the limit of the (local) population density. For constants $r$, $D$, and $K$ (the so-called \emph{isotropic} FKPP equation), one-dimensional solutions evolve to a propagating wavefront which travels at a speed of 
\begin{equation}
	v = 2\sqrt{Dr},
	\label{eq:fkpp-wavespeed}
\end{equation}
if $N(t=0)$ has compact support \cite{kolmogorov_etude_1937}. In two dimensions, this is true provided the curvature of the wavefront is small, such as when the distance of the wavefront from its source is large \cite{murray_mathematical_2002}.  \\

\begin{figure}[!ht]
        \centering 
        \includegraphics[width=\linewidth]{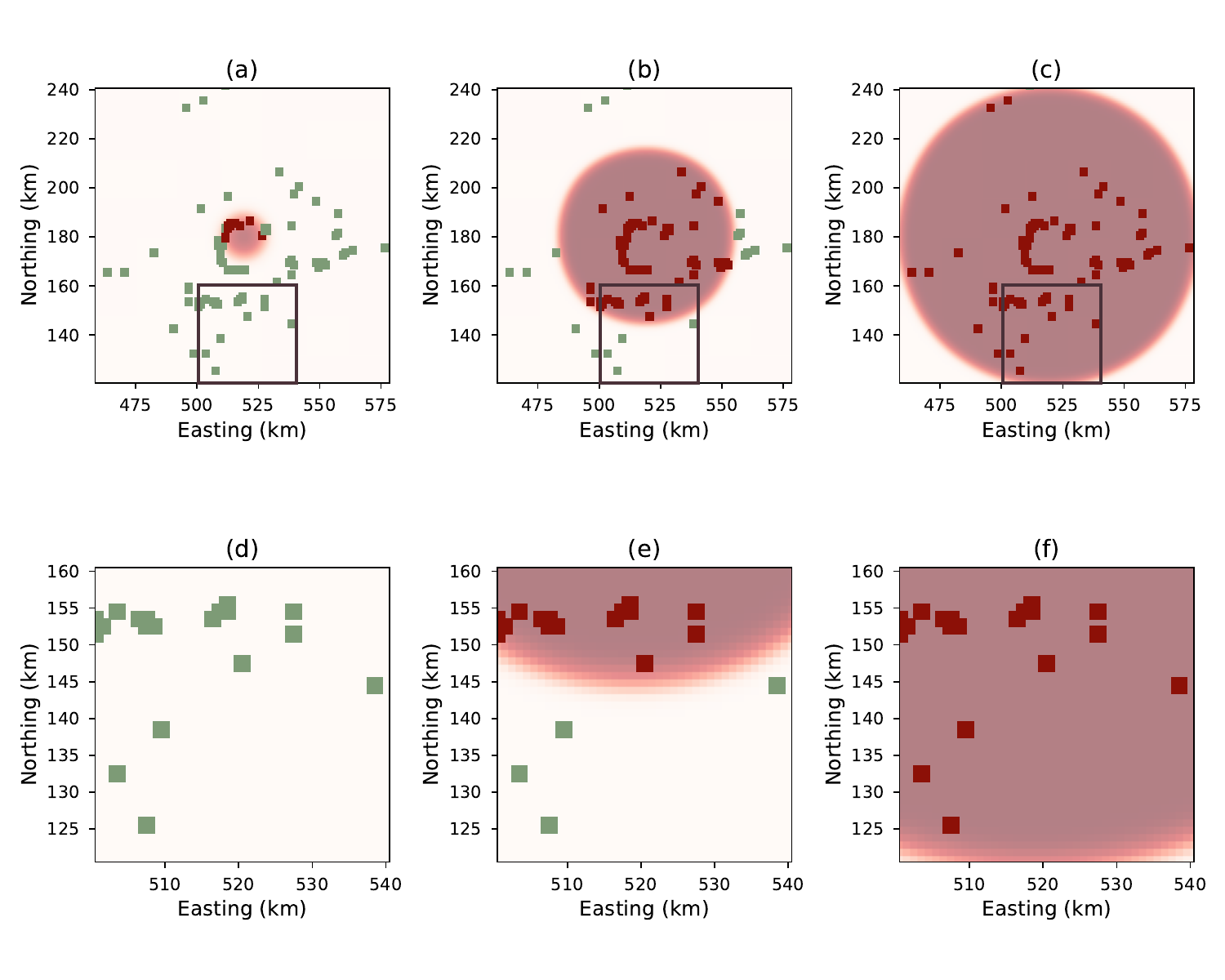}
        \caption{A prototype wavefront solution to Equation~\eqref{eq:fkpp} with $r(t)$ given by Equation~\eqref{eq:fkpp2} with $r_1 = 0.35$, $r_2 = 6.21$, and $t_c = 2014.17$ (shown in pink-red). This solution evolves from the initial distribution given in Section~\ref{sec:data}. Upper row: panel (a) shows $t=2014$, panel (b) shows $t=2018$, and panel (c) shows $t=2022$. Lower row: panels (d), (e), and (f) show close ups of the front at $t= 2014$, $2018$, and $2022$, respectively. The region depicted in panels (d) to (f) is shown as square outlines in panels (a) to (c). Red squares indicate sites that are infested at the corresponding time $t$ (according to this prototype) and green squares indicate sites that are not infested.}
    \label{fig:prototype}	
    \end{figure}

Here, we adopt units of years (yr) for $t$ and units of kilometres (km) for (each component of) $\mathbf{x}$. We assume the spatiotemporal distribution of the pest is described by the population field $N$ and evolves according to the reaction-diffusion equation defined in Equation~\eqref{eq:fkpp}; therefore, the density field evolves with a clear outward-travelling wavefront (see Figure~\ref{fig:prototype}). For the parameters of Equation~\eqref{eq:fkpp}, we assume a constant value of $K=1$ for the carrying capacity and we assume the growth rate $r$ and diffusivity $D$ are uniform in space, but time-dependent. Realistically, the carrying capacity, growth rate, and diffusivity will all depend on space and time due to a range of factors not accounted for by this model, such as weather, seasonality, and host density and distribution \cite{suprunenko_estimating_2021, csoka_weather-dependent_2018, Battisti2015}. \\

Equation~\eqref{eq:fkpp} has a natural lengthscale $\ell$ satisfying $D = \ell^2 r$ (see e.g. \cite{murray_mathematical_2002}). The natural lengthscale is the OPM dispersal distance, so we set $\ell = 0.5$km, given in \cite{sands_population_2017} as the expected distance between a new nest and its parent nest. The wavespeed $v$ in Equation~\eqref{eq:fkpp-wavespeed} can be interpreted as the expansion rate of the population and has units of km/year. Since $\ell=0.5$, we have $v = r$ km/year (with $r>0$, assumed implicitly in what follows). The specific forms we adopt for the growth rate $r$ are given in Section~\ref{sec:proposedmodels}. \\ 

Since analytic solutions to Equation~\eqref{eq:fkpp} are generally intractable, we adopt a numerical approach. For given growth rate $r$, we discretize Equation~\eqref{eq:fkpp} using $6^\text{th}$-order central finite differences with respect to the grid $\Omega$ (defined in Section~\ref{sec:data}), so that each ODE describes the time-evolution of OPM at one of the locations $\mathbf{x}_{ij} = (e_j, n_i) \in \Omega$. We write $N^p = N(\cdot, t_p)$ for the state at time $t_p = t_0 + p\Delta t$. Starting at $p=0$, the system is advanced from $N^p$ to $N^{p+1}$ by a step $\Delta t$ according to a low-storage Runge-Kutta scheme \cite{williamson_low-storage_1980}; we chose $\Delta t = 0.0125$ following convergence tests. We adopt the symmetric $3^\text{rd}$-order scheme defined in \cite{dobler_pencil_2010}, which is parametrised by $a = [0, -2/3, -1]$, $b = [1/3, 1, 1/2]$, and $c = [0, 1/3, 2/3]$ (see \cite{williamson_low-storage_1980} for notation). \\ 

At each time $t_p$, we apply a binary mask to the density field, i.e. $N^* = (N > N_T)$ such that $N_{ij}^* \in \{0,1\}$, where $N_T$ is the arrival threshold. Locations $\mathbf{x}_{ij}$ for which $N^*_{ij} = (N_{ij} > N_T) = 1$ are said to be infested with OPM at time $t_p$. We find all locations $\mathbf{x}_{ij}$  for which $N^*_{i,j} = 1$ \emph{for the first time} (i.e. $N^*_{ij}(t_p) = 1$ and $N^*_{ij}(t) = 0 $ for $t < t_p$) and identify these $\mathbf{x}_{ij}$ with the corresponding observed sites $\mathbf{s}_k$. We thus call $t_p$ the \emph{simulated arrival time} of OPM at these locations (see Section~\ref{sec:bayesianinference}). A schematic of this transition is shown (up close) in the second row of Figure~\ref{fig:prototype}. Here, we adopt $N_T = 0.5$. Note that the adopted lengthscale $\ell = 0.5$ ensures the wavefront is sharp, therefore the arrival time is essentially independent of this value. \\

\subsection{Proposed expansion rate(s)\label{sec:proposedmodels}}
We consider two models of OPM spread, each having the form of the reaction-diffusion equation described in Equation~\eqref{eq:fkpp}. We denote by $M_1$ a model of the form Equation~\eqref{eq:fkpp} with 
\begin{equation}
    r(t) = r, 
    \label{eq:fkpp1}
\end{equation}
for some real value $r>0$. The parametrisation of $M_1$ is $\theta_1 = (r)$, where we adopt vector notation for consistency (see definition of $\theta_2$ below). This \emph{monophasic} model $M_1$ encodes the assumption that the spread of OPM occurs at a constant rate $r$ throughout the period $2006$ to $2023$. The second model $M_2$ is of the form Equation~\eqref{eq:fkpp} with 
\begin{equation}
    r(t) = \begin{cases} r_1 & \text{if}\ t \le t_c, \\ r_2 & \text{otherwise,}\end{cases}
    \label{eq:fkpp2}
\end{equation}
where $r_1,r_2>0$ are real values and the transition time $t_c$ satisfies $2006 \le t_c \le 2023$. Hence, $M_2$ is parametrised by $\theta_2 = (r_1, r_2, t_c)$. This \emph{biphasic} model $M_2$ encodes the assumption that the spread of OPM occurs at a non-constant rate; at a rate of $r_1$ from $2006$ up to and including $t_c$ and at a rate $r_2$ from $t_c$ to $2023$. We provide snapshots of a prototypical numerical solution to this model in Figure~\ref{fig:prototype}. We refer to these models as Model 1 and Model 2, respectively. \\
								
\subsection{Parameter inference}\label{sec:bayesianinference}
Here, we outline the adopted Bayesian approach to parameter inference for each of the proposed models (see Section~\ref{sec:proposedmodels}). For further details, we refer the reader to Appendix~\ref{sec:app_bayesianinference}. \\

We assume that the observed data $\mathcal{D}$ are generated by the proposed reaction-diffusion model(s) subject to spatially-independent Gaussian errors. This captures expected local deviations from the idealized, global model - see e.g. \cite{baggaley_bayesian_2012}. Specifically, we assume that
\begin{equation}\label{eq:obs_model}
t_k \sim N\left(\tau(\mathbf{s}_k | \theta),\sigma^2\right),\ \ k=1,\dots,n,
\end{equation}
where $\sigma$ is the spatially homogeneous standard deviation, which we term the \emph{observation noise parameter}. These statistical assumptions permit a tractable Gaussian likelihood function $\pi(\mathcal{D} | \theta,\sigma)$. Upon ascribing a prior density $\pi(\theta,\sigma)$ to the unknown model parameters, inference proceeds via the posterior distribution $\pi(\theta, \sigma | \mathcal{D})$, given by Bayes' theorem as 
\begin{equation}\label{eq:posterior}
 \pi(\theta,\sigma|\mathcal{D}) \propto  \pi(\theta,\sigma)
 \pi(\mathcal{D} | \theta,\sigma). 
\end{equation}

Note that we have suppressed the dependence of the posterior on the particular model of interest ($M_1$ or $M_2$ and parametrisation $\theta_1$ and $\theta_2$) for notational simplicity. Our initial beliefs about likely values for $\theta$ and $\sigma$ are encoded by the prior density $\pi(\theta,\sigma)$. We adopt an independent prior specification and assume log-normal $\mathrm{LogN}(2, 0.29^2)$ distributions for each of the growth components, $r$ in $M_1$ and $r_1,r_2$ in $M_2$. This specification corresponds to parameter values that are consistent with previous expansion rate estimates \cite{suprunenko_estimating_2021, townsend_oak_2013, groenen_historical_2012} and reduces bias toward a mono- or bi-phasic model. For the transition time component $t_c$, we also adopt a log-normal distribution with median chosen to be the average of $2006$ and $2023$; we take $t_c \sim \text{LN}(\log(2014.5), 0.02^2)$. Finally, we use an inverse Gamma distribution for the (square of the) observation noise parameter, so that $\sigma^2 \sim \text{IG}(4.3,10)$. The variances of our priors are chosen so that the corresponding distributions are weakly-informative.  \\

The posterior in Equation~\eqref{eq:posterior} is unavailable in closed form, necessitating the use of sampling approaches to inference, such as 
Markov chain Monte Carlo (MCMC, see e.g. \cite{gamerman_markov_2006}). Given the natural parameter blocks of $\theta$ and $\sigma$, we adopt a Gibbs sampler (see e.g. \cite{geman_stochastic_1984}), which targets Equation~\eqref{eq:posterior} by alternating between draws of the full conditional distributions of $\theta$ and $\sigma$. The observation model (Equation~\eqref{eq:obs_model}) and inverse Gamma prior permit realizations of $\sigma$ to be sampled directly (as in \cite{baggaley_bayesian_2012}), however the full conditional distribution for $\theta$ is intractable. We therefore adopt a random-walk Metropolis-Hastings step to sample from the corresponding distribution. The resulting \emph{Metropolis-within-Gibbs} step for $\theta$ is constructed by generating proposals for each component of $\theta$ via a symmetric random walk with normal innovations on a logarithmic scale, so that all parameters are non-negative/biologically-feasible. A proposal is accepted as the next value in the chain with a probability that ensures the invariant distribution of the Markov chain simulated by the Gibbs sampler is precisely the target distribution \cite{gamerman_markov_2006}. \\

The innovation variance, given by $\Sigma$, is the strictly positive-definite covariance matrix of the proposal distribution used to update $\theta$. The choice of $\Sigma$ influences mixing of the Markov chain \cite{baggaley_bayesian_2012}. An adaptive Metropolis-Hastings step can improve mixing, wherein $\Sigma$ is updated at fixed intervals \cite{haario_adaptive_2001}. We adopt the approach in \cite{haario_adaptive_2001}, instantiating the Gibbs sampler with an arbitrary initial innovation matrix $\Sigma = \Sigma^0$ (we use the diagonal of the prior covariance matrix). For the first $p$ samples, we update the innovation matrix at every $l\text{th}$ sample using the current covariance of the corresponding chain and a heuristic taken from \cite{Roberts2001}, where $0 < l < p \ll m_T$ and $m_T$ is the total number of samples obtained. \\ 

The MCMC scheme (Gibbs sampler) described above is iterated until $m_T$ samples are obtained. We discard the initial $m_B \ge p$ samples as burn-in, and informally assess convergence of the remaining $m=m_{T}-m_{B}$ samples via visual diagnostics such as trace plots \cite{gamerman_markov_2006}. Additional scaling of the proposal variance to optimise acceptance rates can be carried out post hoc, if desired \cite{schmon_optimal_2022}. See Appendix~\ref{sec:app_bayesianinference} for more details.\\

\subsection{Model validation}\label{sec:modelval}
We run the MCMC scheme for each of the models $M_i$ described above. To compare the models, we use the \emph{deviance information criterion} (DIC), which measures relative quality of models \cite{spiegelhalter_bayesian_2002}. For model $M_i$ (parametrised by $\theta_i$), the DIC is given by
\begin{equation}
\text{DIC}_i = 2\overline{D(\theta_i)} - D(\bar{\theta}_i)\label{eq:DIC},
\end{equation}
where $D(\theta_i) = -2\log(\pi(\mathcal{D} | \theta_i))$ is the deviance of $\theta_i$; this criterion quantifies goodness of fit while penalising over-fitting. Models with a smaller DIC should be preferred to models with a larger DIC. We can readily compute $\bar{\theta}_i$ and $\overline{D(\theta_i)}$ using the samples of $\theta$, obtained from the (final) MCMC run. Finally, we assess goodness of fit of the model with lowest DIC using \emph{posterior predictive sampling}. Let $\tilde{t}_k$ denote a first arrival time at site $\mathbf{s}_k$. The predictive density is given by
\begin{equation}
    \pi\left(\tilde{t}_k | \mathcal{D}\right) = \iint f\left(\tilde{t}_k | \tau\left(\mathbf{s}_k | \theta\right), \sigma\right)\pi\left(\theta,\sigma | \mathcal{D}\right) \ \mathrm{d}\theta\mathrm{d}\sigma,
\end{equation}
where $f$ denotes the Gaussian density with mean $\tau(\mathbf{s}_k | \theta)$ and variance $\sigma^2$. Hence, given samples $\left\{\left(\theta^{(j)}, \sigma^{(j)}\right)\right\}_{j=1}^m$ from $\pi\left(\theta,\sigma | \mathcal{D}\right)$, we generate samples $\tilde{t}_k^{\ (j)}$ from $\pi\left(\tilde{t}_k | \mathcal{D}\right)$ by drawing from $\tilde{t}_k^{\ (j)} \sim N\left(\tau\left(\mathbf{s}_k | \theta^{(j)}\right), \left(\sigma^{(j)}\right)^2\right)$ for $j=1,\dots,m$. Thus, under the preferred model, we generate $m=10^3$ predicted observations for each observed site. We select nine survey years spaced across the whole survey period. From each of these years, we calculate the distance (from the centre $\mathbf{x}^c$) of all sites observed in that year and select the site with distance closest to the median distance. These nine selected test sites (given in Table~\ref{tbl:obssites}) provide reasonable spatial coverage of the observed sites. \\

\begin{center}
	\begin{table}[!ht]
        \centering
		\begin{tabular}{c c  c c}
		\hline 
        \rowcolor{myBrown!25} Year		&	Northing (km)		&	Easting (km)	&	$d$ (km)	\\ 
		\hline 
		$2008$	& 	$177.5$	&	$520.5$	&	$3.61$	\\
        \cdashline{1-4}
		$2010$	&	$174.5$	&	$518.5$	&	$6.00$ 	\\
        \cdashline{1-4}
		$2012$	&	$173.5$	&	$522.5$	&	$8.06$	\\
        \cdashline{1-4}
		$2014$	&	$178.5$	&	$509.5$	&	$9.22$	\\
        \cdashline{1-4}
		$2016$	&	$181.5$	&	$509.5$	&	$9.06$	\\
        \cdashline{1-4}
		$2018$	&	$161.5$	&	$532.5$ 	&	$23.60$	\\
        \cdashline{1-4}
		$2020$	&	$182.5$	&	$476.5$	&	$42.05$	\\
        \cdashline{1-4}
		$2022$	&	$187.5$	&	$583.5$	&	$65.38$	\\
        \cdashline{1-4}
		$2023$	&	$195.5$	&	$595.5$	&	$78.45$	\\
        \cdashline{1-4}
		\hline
		\end{tabular}
		\caption{Table of nine observed sites selected to validate the best performing proposed model. The fourth column $d$ is the distance from $\mathbf{x}^c = (518.5, 180.5)$ to that point in kilometres (km); $\mathbf{x}^c$ is the centroid of the initial states (see Section~\ref{sec:data}).}
		\label{tbl:obssites}
	\end{table}
\end{center}

\section{Results}\label{sec:results}
For each model, we run the adaptive MCMC scheme to obtain $5.0 \times 10^4$ samples; during these runs, adaptive tuning was applied every $100$ samples until the ${(10^4)}^\text{th}$ sample was obtained. \\ 

\begin{figure}[!ht]
        \centering
        \includegraphics[width=\textwidth]{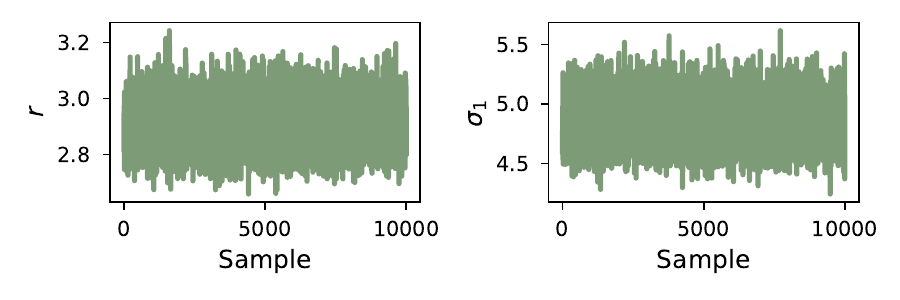}
        \caption{Model 1. Trace plots based on the final MCMC run for each parameter chain from the output of the MCMC scheme.}
        \label{fig:tracesM1}
    \end{figure}

\begin{figure}[!ht]
    \centering
    \includegraphics[width=\textwidth]{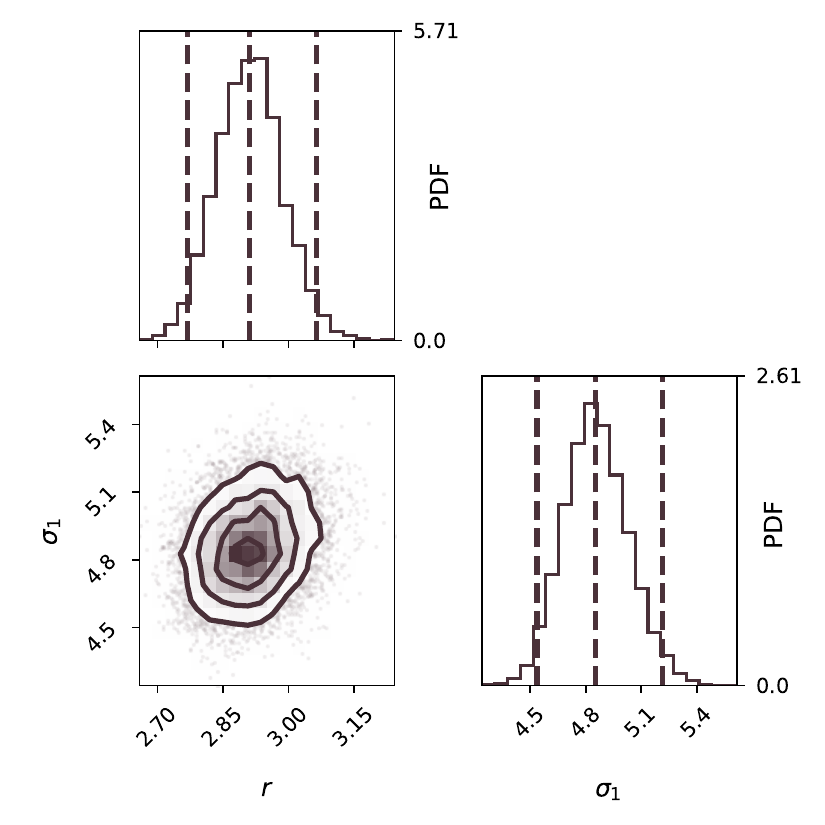}
    \caption{Model 1. Posterior distributions (histograms) and kernel density estimate of the joint posterior (contour plot), using $10$K samples of the MCMC scheme. (Dashed) lines indicate the posterior mode and $95\%$ credible interval.}
    \label{fig:posteriorsM1}
\end{figure}

Recall that Model 1 is parametrised by $\theta_1 = (r)$ and $\sigma_1$. We initialise the MCMC scheme with values $\theta_{1,0} = 4.0$ and $\sigma_{1,0} = 2.0$, chosen arbitrarily. After burn-in and thinning the sample output to reduce auto-correlation, we obtain $m_1=10^4$ samples from the posterior distribution of the parameters of Model 1. This output is summarised by Figures~\ref{fig:tracesM1} and \ref{fig:posteriorsM1}, with the former suggesting satisfactory convergence of the MCMC scheme. In Table~\ref{tbl:results}, we provide a summary of the inferred parameters for each model. For Model 1, with a constant expansion rate, we obtain a posterior mode of $\hat{r} = 2.91$ with $95\%$ (equi-tailed) credible interval (CI) $(2.77, 3.06)$ for the growth parameter $r$. For the observed noise parameter $\sigma$, the posterior mode is $\hat{\sigma}_1 = 4.85$ with $95\%$ CI $(4.54, 5.21)$. The DIC reported for this model is $\text{DIC}_1 = -1728.07$ (2dp). \\

Model 2 is parametrised by $\theta_2 = (r_1, r_2, t_c)$ and $\sigma_2$. We initialise the MCMC scheme with $\theta_{2,0} = (4.0, 4.0, 2010.0)$ and $\sigma_{2,0} = 2.0$; initial values for the parameters in common with Model 1 are chosen as above and the remaining values are chosen arbitrarily. Model 2 -- with a higher dimensional parameter space -- required a larger thinning factor to obtain near-independent samples. Thus, we obtain $m_2 = 2.2 \times 10^3$ samples from the posterior of Model 2; see Figures~\ref{fig:tracesM2} and \ref{fig:posteriorsM2}, and Table~\ref{tbl:results}. For this model of a non-constant expansion rate, we obtain posterior modes of $\hat{r}_1 = 0.35$ (CI: $(0.32,0.39)$), $\hat{r}_2 = 6.21$ (CI: $(5.88,6.62)$), $\hat{t}_c = 2014.17$ (CI: $(2013.83, 2014.48)$), and $\hat{\sigma}_2 = 0.35$ (CI: $(0.32,0.39)$). The DIC is $\text{DIC}_2 = -830.33$ (2dp). \\

\begin{figure}[!ht]
    \centering
    \includegraphics[width=\textwidth]{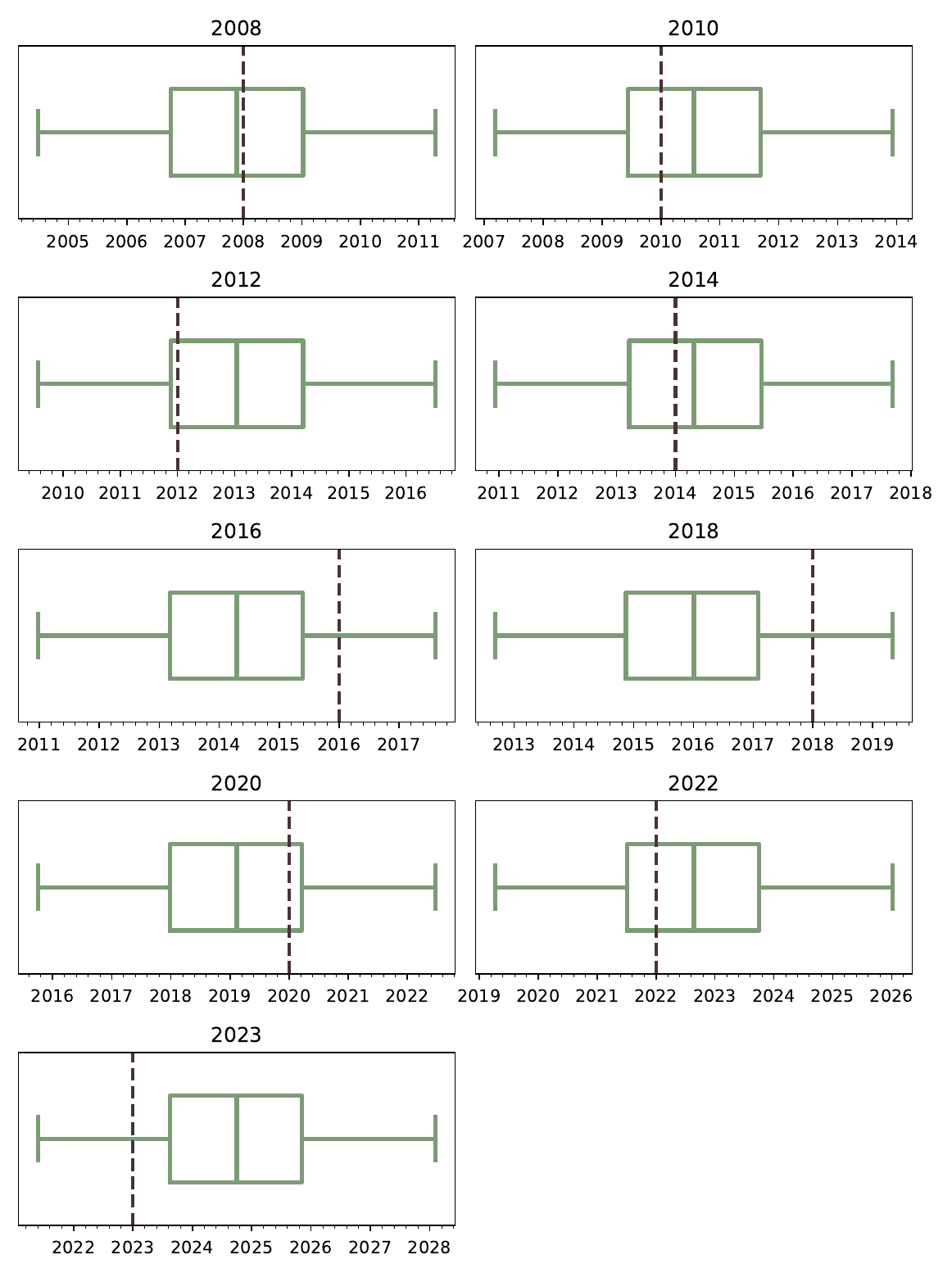} 
    \caption{Box plots summarising the predictive distribution at nine observed sites (test sites) (see Table~\ref{tbl:obssites}). Whiskers indicate $\pm 1.5$IQR. Observations are indicated by a vertical dashed line, with the corresponding observed year above the axis.}
    \label{fig:boxplots}
\end{figure}

\begin{figure}[!ht]
    \centering
        \includegraphics[width=\textwidth]{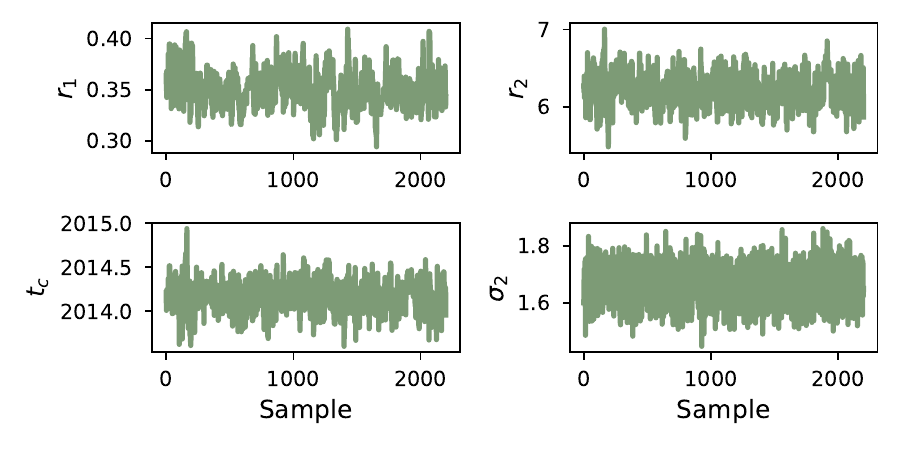}
        \caption{Model 2. Trace plots based on the final MCMC run for each parameter chain from the output of the MCMC scheme.}
        \label{fig:tracesM2}
    \end{figure}

\begin{figure}[!ht]
        \centering
        \includegraphics[width=\textwidth]{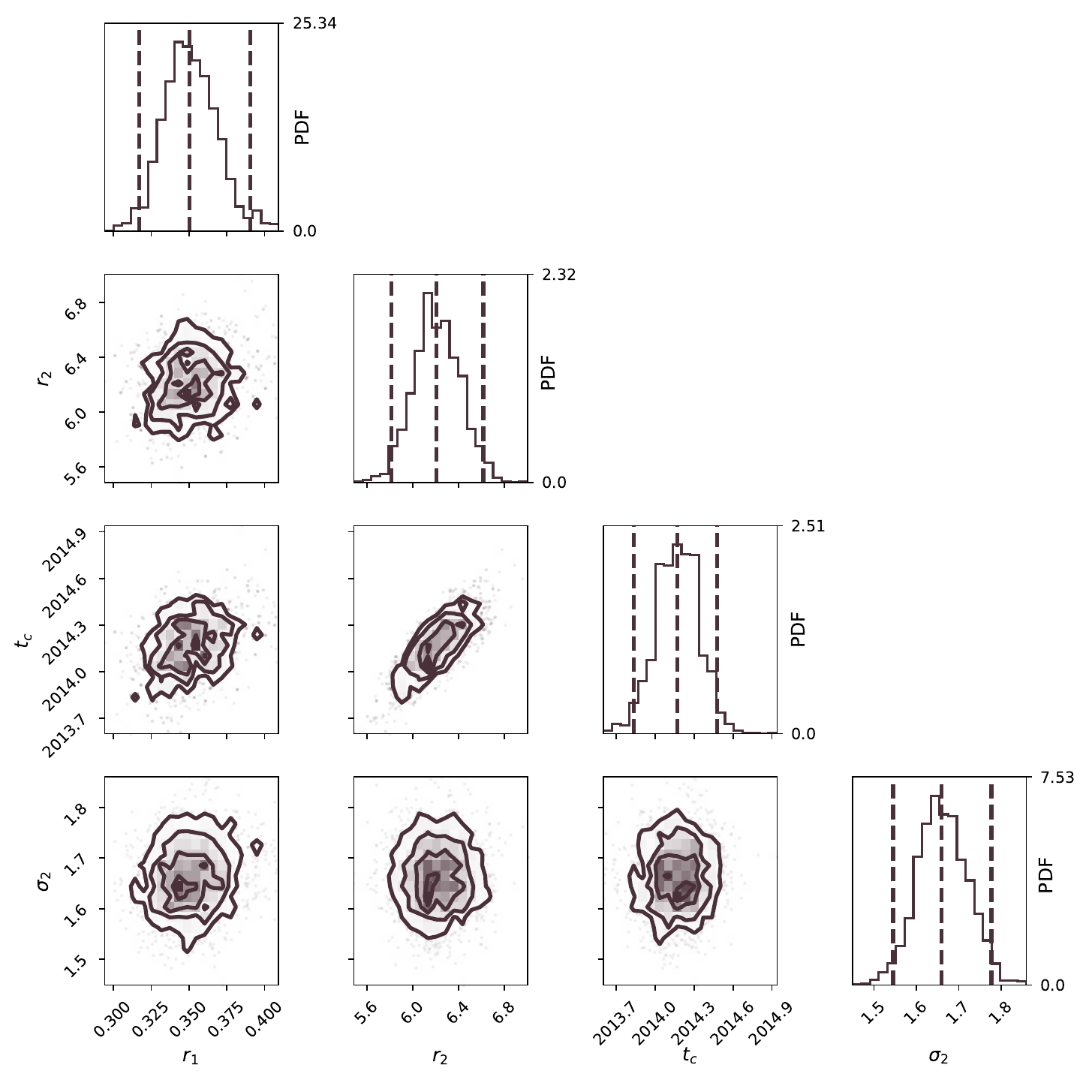} 
        \caption{Model 2. Posterior distributions (histograms) and kernel density estimates of the pairwise marginal posteriors (contour plots), using $2.2$K samples of the MCMC scheme. (Dashed) lines indicate the posterior mode and $95\%$ credible interval.}
        \label{fig:posteriorsM2}
    \end{figure}

The values for the DICs of Models 1 and 2 (Table~\ref{tbl:results}) indicate that Model 2 is the more appropriate choice for the observed data. This suggests that a biphasic expansion model is more appropriate for describing the spread of OPM than a monophasic model. Using the posterior of Model 2, we generated $10^4$ posterior predictive samples for selected observed sites (see Section~\ref{sec:data}) and present box plots summarising the posterior predictive distribution at each site in Figure~\ref{fig:boxplots}. It is clear that the model is able to produce predictions that are consistent with the observations, with most data points lying within the interquartile range (IQR) of the corresponding predictive distribution, and all within a factor of 1.5 of the IQR.   \\
					
\section{Discussion}\label{sec:discussion}
In this section, we begin with comments pertaining to the interpretation of these results and model validity; in Section~\ref{sec:discussion_opm}, we discuss the implications of these results for the spread of oak processionary moth in the UK, finishing with ecological considerations. In Section~\ref{sec:discussion_approach}, we comment on the adopted modelling approach. Finally, in Section~\ref{sec:discussion_furtherwork}, we highlight some potential directions for future work. \\

\subsection{The spread of oak processionary moth in the UK}\label{sec:discussion_opm}

The values for the DICs of Models 1 and 2 (Table~\ref{tbl:results}) indicate that Model 2 is the more appropriate choice for these observed data. This suggests that the expansion of oak processionary moth in the UK from 2006 to 2023 occurred at a non-constant rate. This conclusion aligns with previous results in the literature \cite{suprunenko_estimating_2021}.\\
	
The posterior for Model 1 suggests an expansion rate of $\hat{r} = 2.91$km/year (CI: $(2.77, 3.06)$); this is lower than estimates reported in the literature \cite{suprunenko_estimating_2021, stigter_thaumetopoea_1997,  townsend_oak_2013, groenen_historical_2012}. However, we use survey data for the full period $2006$ to $2023$; no previous work determines expansion rates using data including the period $2019$ to $2023$. These results could suggest that the rate of expansion from $2019$ to $2023$ was lower than the expansion rate from $2006$ to $2019$. Additionally, we note that the posterior indicates a modal observed noise parameter of $4.85$ years (CI: $(4.54, 5.21)$); this is fairly large with respect to the $18$-year period these data cover and suggests the constant-rate model does not accurately estimate the arrival of OPM at the observed sites (on average). This could be due to the prescribed initial state for the distribution of OPM in $2006$ (see Section~\ref{sec:methods}); all observed sites from $2007$ to $2013$ are located where the corresponding initial state $N^0$ is non-zero (see Section~\ref{sec:model}), thus our model may obtain earlier arrival times for these sites than the true values, irrespective of model parameter choices. \\

The posterior for Model 2 suggests that the spread occurs in two temporal phases at (posterior mode) rates of $\hat{r}_1 = 0.35$ and $\hat{r}_2 = 6.21$, with (posterior mode) phase transition time $\hat{t}_c =  2014.17$. The posterior mode of the observed noise parameter is $\hat{\sigma}_2 = 1.66$. Comparison with the posterior mode for Model 1 ($\hat{\sigma}_1 = 4.85$) indicates that a biphasic model is more suitable for describing these data than a monophasic model. Since the MCMC for Model 2 is instantiated with the same initial growth rates and corresponding priors, matching those of Model 1, the difference between the phase 1 and phase 2 growth rate posteriors gives a strong indication that OPM expansion occurs at a non-constant rate. This conclusion is further supported by the DICs. Box plots of the posterior predictive observations are provided in Figure~\ref{fig:boxplots}. With the exception of $2016$, $2018$, and $2023$, all true observations lie within the interquartile range of the posterior predictions for that year. This further supports the applicability of a non-constant expansion rate to quantify the spread of OPM in the UK. The three erroneous predictions could be explained by our removal of points via the convex hull (see Section~\ref{sec:data}); in some years (such as 2014), the hull's edge is smoother (i.e. made up of more observations), providing a more accurate estimate of the wavefront position than in $2023$ (see Figure~\ref{fig:convexlayers}). \\

Our evidence of biphasic expansion aligns with previous results, which also indicated two phases of expansion (at rates $1.66$ and $6.17$, respectively) with transition time $2014$ \cite{suprunenko_estimating_2021}. We note that the weighted average of $\hat{r}_1$ and $\hat{r}_2$ ($3.28$) is reasonably close in value to the Model 1 growth rate ($2.91$), and closer still to the mean of the previous biphasic estimates ($3.27$) \cite{suprunenko_estimating_2021}. Since the latter estimate covers $2006$ to $2019$, this could suggest that the expansion rate of OPM has continued to increase from $2019$ to $2023$. This is further supported by the increase in the phase 2 growth rate ($\hat{r}_2 = 6.21$ vs $6.17$). Nevertheless, we note that $6.17$ is within the 95CI of $r_2$, so it is possible the rate has stayed the same or decreased. In addition, methodological differences could explain these discrepancies.\\

Ecologically, a constant rate of expansion could be deemed unrealistic; the spread of a pest is likely influenced by landscape and climatic variability \cite{blaser_oak_2022, csoka_weather-dependent_2018, sands_population_2017}. These factors could be time-dependent; seasonality results in cyclic temperature trends and consumption of woodland by society (e.g. via the forestry industry) varies with demand and thus possibly with time. Additionally, it was noted in \cite{Memmott2005} that some invasive populations can initially decline before slowly increasing over several years. This could manifest in terms of an initial slower phase of expansion, in which a declining population results in fewer new $1\text{km}^2$ areas where the pest is observed. A second phase of expansion, characterised by a higher expansion rate, could follow due to the adaptive evolution of the pest to its new environment \cite{sands_population_2017, Lee2002}, which is expected to occur over several generations \cite{Lee2002}. Therefore we would expect to capture the biphasic expansion of OPM over multiple years, as indicated by the proposed biphasic model. Alternatively, the apparent biphasic spread of OPM could be the result of spatial variability, in landscape and climate, that are known to impact the spread of a pest \cite{blaser_oak_2022, csoka_weather-dependent_2018, sands_population_2017}. Furthermore, we model dispersal as diffusion; spatially-explicit diffusion can be incorrectly interpreted as time-dependence \cite{bentz_time-dependent_2001}. \\

\subsection{Modelling approach\label{sec:discussion_approach}}
Previous estimates of the expansion rate of OPM in the UK were most recently obtained using a maximum distance method \cite{suprunenko_estimating_2021}. With our reaction-diffusion approach, we can obtain an estimate of the rate directly from our model parameters. This offers benefits; this framework can readily estimate future expansion rates. However, the equation for the expansion rate (Equation~\eqref{eq:fkpp-wavespeed}) is obtained analytically, hence it will be susceptible to error introduced by the choice of numerical integration method. Fortunately, testing indicates acceptable convergence of the numerical scheme for our model, spatial-,  and temporal-scales. \\

Our spatially-explicit modelling approach can provide an estimate of the arrival of OPM at every $1\text{km}$ area in a $256\text{km}^2$ spatial domain and can be readily extended to larger domain sizes, provided adequate computational resources are available. Our reaction-diffusion equation describes the evolution of the population density field; we apply a binary mask to this to facilitate comparison of our model with observed presence/absence data. Additionally, this could facilitate direct comparison of our approach with similar (compartmental) approaches \cite{wadkin_inference_2022, suprunenko_predicting_2025}. Furthermore, the population density field evolves with a clear outward-travelling wavefront (as in Figure~\ref{fig:prototype}). This captures the spatial trend of first arrival of the pest without incurring the computational cost of simulating a large number of individuals, for example. The use of a binary mask also allows our model to somewhat account for potential time lag in observation; time lag is posited to reflect the time needed for populations to reach detectable densities \cite{siegert_dendrochronological_2014}. \\

The adopted form of a reaction-diffusion equation permits calibration of the dynamics equation (Equation~\ref{eq:fkpp}) to modelling OPM by fixing a specific lengthscale $\ell$, taken from available estimates of the expected dispersal distance of a (female) OPM in a single dispersal event \cite{sands_population_2017}. This facilitates application of our framework to other invasive pests, provided suitable spatial and temporal scales can be adopted and a reasonable estimate for $\ell$ obtained. This lengthscale $\ell$ determines the motility of the population with respect to its (local) reproduction, thus results may be highly sensitive to inaccuracies in estimating $\ell$. This could be challenging for problems where the primary goal is to determine risk of infestation in previously uninfested areas. Our adopted lengthscale for OPM is $\ell = 0.5$ \cite{sands_population_2017}; most adult females probably nest between $0$ and $1$km from where they emerge \cite{sands_population_2017}. Other estimates are available, ranging from $5$ to $20$km \cite{sands_population_2017, stigter_thaumetopoea_1997, groenen_historical_2012}, but these estimates have been attributed to extreme female flight capabilities. Therefore, the choice of $\ell = 0.5$ is reasonable, although it may be of interest to consider other lengthscales $\ell \in (0.5,1.0]$. \\

Reaction-diffusion models, such as the one considered here, require specification of several parameters, resulting in analytically intractable posterior distributions. In this work, we adopt a Bayesian approach to parameter inference, via computation of the joint posterior distribution over all parameters of interest, using data consisting of first arrival times at some $414$ locations across the South of England. We assume that the true underlying arrival time is not observed exactly, but subject to Gaussian noise, giving a tractable likelihood function. However, the posterior distribution is intractable, necessitating the use of sampling-based approaches such as MCMC \cite{gamerman_markov_2006}. In particular, these are routinely employed to infer parameters for models of epidemics \cite{wadkin_quantifying_2023, golightly_accelerating_2023, wadkin_inference_2022} and collective behaviour \cite{baggaley_bayesian_2012, walton_bayesian_2021}; the adopted scheme requires the likelihood to be known only up to proportionality. Specifically, we adopt a Metropolis-within-Gibbs scheme; such approaches are popular among ecological population modellers \cite{ wadkin_quantifying_2023, golightly_accelerating_2023, wadkin_inference_2022, baggaley_bayesian_2012, haario_adaptive_2001, walton_bayesian_2021}. We inferred parameters of the models using an adaptive random-walk Metropolis-within-Gibbs approach to generate posterior samples. Model 2 -- with a four-dimensional parameter space -- had to be run for longer than Model 1 to obtain a feasible number of near-independent samples from the posterior. However, we note no significant issues with convergence. For model comparison, we use the deviance information criterion to assess the relative quality of the two competing models. This metric penalizes models based on goodness of fit and the effective number of parameters, and is routinely applied when fitting models in the Bayesian paradigm \cite{spiegelhalter_bayesian_2002}. Other metrics can be computed, such as the (frequentist) Akaike information criterion (AIC), which requires maximising the likelihood to obtain the maximum likelihood estimate, rather than estimating this directly from the sample average. As an exercise, we calculated the AIC and found this also indicated Model 2 as the preferred model. \\

   \begin{center}
        \begin{table}[!ht]
            \centering 
            \begin{tabular}{c c c c c c c c } 
            \hline
             \rowcolor{myBrown!25}Model &	Expansion rate & $\hat{\cdot}$	& Mean 	& 95\% CI  &  DIC\\
            \hline
            \multirow{2}{*}{1}		&\multirow{2}{*}{Constant} & $r$ 		& $2.91$		& $(2.77, 3.06)$	& \multirow{2}{*}{$-1728.07$}				\\
                                & & $\sigma_1$	& $4.85$		& $(4.54, 5.21)$	&							\\
            \cdashline{1-8}
            \multirow{4}{*}{2}		&\multirow{4}{*}{Non-constant}& $r_1$ & $0.35$		& $(0.32,0.39)$	& \multirow{4}{*}{$-830.33$}  				\\
                                && $r_2$		& $6.21$		& $(5.88, 6.62)$	&	 							\\
                                && $t_c$		& $2014.17$ 	& $(2013.83, 2014.48)$	&	 							\\
                                && $\sigma_2$	& $1.66$  	& $(1.55, 1.78)$	&	 							\\ 	
            \hline 
            \end{tabular}
            \caption{Table of results providing, for each of the two proposed models, the calculated DIC as well as the mean and $95\%$ equi-tailed credible interval for each of the model parameters.}
            \label{tbl:results}
        \end{table}
    \end{center}

\subsection{Further Work}\label{sec:discussion_furtherwork}
Our work, in particular, could indicate that the spread of OPM in the UK occurred in two phases, but further work would illuminate whether there are more phases of expansion. This apparent time-dependence could be misleading; it is known that spatial dependence of the diffusivity $D$ can be misinterpreted as time-dependence \cite{bentz_time-dependent_2001}. Furthermore, dispersal capabilities and population resource constraints are highly likely to depend on environmental factors, such as temperature, precipitation, and host distribution \cite{blaser_oak_2022, csoka_weather-dependent_2018, sands_population_2017}. Consequently, we believe that environmental factors -- in particular dependence of spread on host distribution -- are the likely reason for the apparent biphasic expansion of OPM in the UK. In this work, we adopt a reductionist approach \cite{NartalloKaluarachchi2025}, whereby we omit dependence on landscape and environmental factors in favour of a simpler model. Further study is needed to understand how host density affects OPM spread and how this is best incorporated into a reaction-diffusion model. We note that recent work used a coupled compartmental-dispersal model to predict the effect of landscape structure on epidemic invasions of agricultural crops \cite{suprunenko_predicting_2025}. \\


\section*{Acknowledgements} 
We thank Julia Branson from the University of Southampton and Andrew Hoppit 
from Forestry Commission England for useful discussions on the application to OPM, and the former for providing the OPM observational data. We also thank Kaitlyn N. Ries for helpful feedback on the manuscript.

\section*{Funding} 
This work was completed as part of JPM's PhD, supported by the Natural Environment Research Council (NERC) (Grant Number [NE/S007512/1]). LEW acknowledges support from a NERC Knowledge Exchange Fellowship (Grant Number [NE/X000478/1]). 

\section*{Author contributions}
\textbf{Jamie P. McKeown} [conceptualization, methodology, software, validation, formal analysis, investigation, data curation, writing -- original draft, writing -- review and editing, visualization, project administration]; \textbf{Laura E. Wadkin} [conceptualization, methodology, data curation, writing -- review and editing, supervision, project administration, funding acquisition]; \textbf{Nick G. Parker} [conceptualization, methodology, data curation, writing -- review and editing, supervision, project administration, funding acquisition]; \textbf{Andrew Golightly} [methodology, writing -- review and editing]; \textbf{Andrew W. Baggaley} [conceptualization, methodology, data curation, writing -- review and editing, supervision, project administration, funding acquisition]. 

\section*{Data availability}
The observational OPM nest data is available from Forestry Commission, England. Background maps of the UK used in Figures 1 and 2 are derived from oak abundance data in \cite{hill_abundance_2017}, available at \cite{hill_abundance_dataset}. Code and data produced by this work are available at \cite{fkpp_opm_data_code}. 

\def\bibcommenthead{}%
\bibliographystyle{sn-mathphys-num.bst}
\bibliography{references}

\appendix 

\section{Inferring model parameters}\label{sec:app_bayesianinference}
The adopted statistical assumptions (see Section~\ref{sec:bayesianinference}) permit a tractable likelihood function $\pi(\mathcal{D} | \theta,\sigma)$ given, up to proportionality, by
\begin{equation}
\pi(\mathcal{D} | \theta,\sigma) \propto \sigma^{-n} \exp\left[-\frac{1}{2\sigma^2}\sum_{k=1}^n \left\{t_k - \tau(\mathbf{s}_k | \theta)\right\}^2\right]. \label{eq:likelihood}
\end{equation}
The posterior distribution $\pi(\theta, \sigma | \mathcal{D})$ of the parameters given the observed data $\mathcal{D}$ is provided by Bayes' theorem in Equation~\eqref{eq:posterior}. Due to the complex dependence of the numerical solution to \eqref{eq:fkpp} on the model parameters, the posterior is analytically intractable. Bayesian practitioners commonly use Markov chain Monte Carlo (MCMC) methods to sample from posterior distributions that are available up to an unknown constant of proportionality. Here, we adopt a particular MCMC scheme known as a Gibbs sampler \cite{geman_stochastic_1984}. In absence of analytically tractable full conditionals, a Metropolis-Hastings step can be used. Such an approach is termed a \emph{Metropolis-within-Gibbs} scheme. Since the target posterior $\pi(\theta, \sigma | \mathcal{D})$ is analytically intractable, we sample from the full conditionals instead, alternating between draws of $\theta$ and draws of $\sigma^2$ (and $\sigma$) \cite{baggaley_bayesian_2012, geman_stochastic_1984}. This procedure is defined as follows:

\begin{enumerate}
	\item Initialize $\theta^0$ and $\sigma^0$; set $j=1$ 
	\item Draw $\sigma^j \sim \pi(\cdot | \theta^{j-1}, \mathcal{D})$
	\item Draw $\theta^{(j)} \sim \pi(\cdot | \sigma^{j}, \mathcal{D})$
	\item Set $j = j+1$ and return to step 2.
\end{enumerate}	

In step 3, the full conditional for $\theta$ is intractable, hence we adopt a random walk Metropolis-Hastings step to sample from the corresponding distribution. Define $\lambda = \log(\theta)$, where the logarithm is applied component-wise. In step 3 of the Gibbs sampler, we propose a new value $\lambda^* = \lambda + {\omega}$, where ${\omega} \sim N({0}, \Sigma)$, for a strictly positive-definite innovation matrix $\Sigma$ -- the covariance matrix of the proposal distribution. The choice of innovation matrix $\Sigma$ influences the mixing of the Markov chain. Practitioners often adopt an \emph{adaptive} MCMC scheme wherein the innovation matrix $\Sigma$ is updated at fixed intervals to improve mixing \cite{haario_adaptive_2001}. We adopt the adaptive approach of Haario \textit{et al} \cite{haario_adaptive_2001}. The MCMC scheme is instantiated with an arbitrary, strictly positive-definite, diagonal initial innovation matrix $\Sigma = \Sigma^0$ (we use the diagonal of the prior covariance matrix with zero covariances). We select a \emph{tuning period} $p \ll m$ and a \emph{tuning interval} $l \ll p$ where, for numerical convenience, we assume $p$ is divisible by $l$. Let $\theta^{(j)}$ denote the $j$th value in the chain. For the first $p$ samples, we update the innovation matrix at every $l^\text{th}$ sample. Hence the innovation matrix for the $j^\text{th}$ step of the MCMC scheme is given by 
\begin{equation}
\Sigma^j = \begin{cases}
			\Sigma^0 & \text{if}\ j < l ,\\
			\Sigma^{j-1} & \text{if}\ j/l \notin \mathbb{Z}\ \text{and}\ j < p,\\
			s_{d_k} \mathrm{cov}(\theta^0, \dots, \theta^{j-1}) & \text{if}\ j/l \in \mathbb{Z}\ \text{and}\ j < p,\\
			\Sigma^p & \text{otherwise,} \\	
\end{cases}
\label{eq:adaptivetuning}
\end{equation}
where the scaling parameter $s_{d_k} = (2.38^2)/{d_k}$ is a heuristic taken from \cite{Roberts2001} and $d_k$ is the dimension of the parameter space of model $M_k$. For a justification of this scheme, see \cite{haario_adaptive_2001}. \\ 
\end{document}